\documentclass[english,manuscript]{pnastwo}

\usepackage[dvips]{graphicx}
\usepackage{pnastwoF}
\usepackage{amssymb,amsfonts,amsmath}
\usepackage{amsmath}
\usepackage{amssymb}
\usepackage{esint}

\DeclareMathOperator{\trace}{Tr}

\makeatother

\usepackage{babel}

\newcommand{\figwidth}{3.375in}

\url{www.pnas.org/cgi/doi/10.1073/pnas.0709640104}
\copyrightyear{2008}
\issuedate{Issue Date}
\volume{Volume}
\issuenumber{Issue Number}

\begin{document}

\title{Quantum State and Process Tomography of Energy Transfer Systems via
Ultrafast Spectroscopy}

\author{Joel Yuen-Zhou\affil{1}{Department of Chemistry and Chemical Biology, Harvard University, Cambridge, MA 02138},
Jacob J. Krich\affil{1},,
Masoud Mohseni\affil{1},,
\and
Alán Aspuru-Guzik\affil{1}{}}

\contributor{Submitted to Proceedings of the National Academy of Sciences
of the United States of America}

\maketitle

\begin{article}
\begin{abstract}
The description of excited state dynamics in energy transfer systems
constitutes a theoretical and experimental challenge in modern chemical
physics. A spectroscopic protocol which systematically characterizes
both coherent and dissipative processes 
of the probed chromophores is desired. Here, we show
that a set of two-color photon-echo experiments performs quantum state
tomography (QST) of the one-exciton manifold of a dimer by reconstructing
its density matrix in real time. This possibility in turn allows for
a complete description of excited state dynamics via quantum process
tomography (QPT). Simulations of a noisy QPT experiment for an inhomogeneously
broadened ensemble of model excitonic dimers show that the protocol
distills rich information about dissipative excitonic dynamics, which
appears nontrivially hidden in the signal monitored in single realizations
of four-wave mixing experiments.

\end{abstract}


\abbreviations{QPT, quantum process tomography; QST, quantum state tomography; PE, photon echo}

\dropcap{E}xcitonic systems and the processes triggered upon their interaction
with electromagnetic radiation are of fundamental physical and chemical
interest \cite{khalil:362,Stone05292009,collinischoles,moran,engelchicago,harel}. In nonlinear
optical spectroscopy (NLOS), a series of ultrafast femtosecond pulses
induces coherent vibrational and electronic dynamics in a molecule
or nanomaterial, and the nonlinear polarization of the excitonic system
is monitored both in the time and frequency domains \cite{mukamel,minhaengbook}.
To interpret these experiments, theoretical modeling has proven essential,
framed within the Liouville space formalism popularized by Mukamel
\cite{mukamel}. Implicit in these calculations is the evolution of
the quantum state of the dissipative system in the form of a density
matrix. The detected polarization contains information of the time
dependent density matrix of the system, although not in the most transparent
way. An important problem is whether these experiments allow quantum
state tomography (QST), that is, the determination of the density
matrix of the probed system at different instants of time \cite{walmsley,cinaprl}.
A more ambitious question is if a complete characterization of the
quantum dynamics of the system can be performed via quantum process
tomography (QPT) \cite{nielsenchuangbook,dcqd}, a protocol that we define
in the next section. In this article, we show that both QST and QPT
are possible for the single-exciton manifold of a coupled dimer of
chromophores with a series of two-color photon echo (PE) experiments.
We also present numerical simulations on a model system and show that
robust QST and QPT is achievable even in the presence of experimental
noise as well as inhomogeneous broadening.


This article provides a conceptual presentation, and interested readers
may find derivations and technical details in the Supporting Information
(SI).

\section{Basic concepts of QPT}

Consider a quantum system which interacts with a bath. We can describe
the full state of the system and the environment at time $T$ by the
density matrix $\rho_{\text{total}}(T)$. The reduced density matrix
of the system is $\rho(T)=\trace_{B}\rho_{\text{total}}(T)$, where
the trace is over the degrees of freedom of the bath. If the initial
state is a product, $\rho_{\text{total}}(0)=\rho(0)\otimes\rho_{B}(0)$
(always with the same initial bath state $\rho_{B}(0)$), then the
evolution of the system may be expressed as a linear transformation
\cite{sudarshan}:

\begin{equation}
\rho_{ab}(T)=\sum_{cd}\chi_{abcd}(T)\rho_{cd}(0).\label{eq:linear transformation}\end{equation}
The central object of this article is the \emph{process matrix} $\chi(T)$,
which is independent of the initial state $\rho(0)$. As opposed to
master equations which are written in differential form, Eq.\ \ref{eq:linear transformation}
can be regarded as an integrated equation of motion for every $T$.
It holds both for Markovian and non-Markovian dynamics of the bath, and it always leads to positive density matrices.
Note that $\chi(T)$ completely characterizes the dynamics of the
system. Preserving Hermiticity, trace, and positivity of $\rho(T)$
imply, respectively, the relations (SI-I)
\begin{eqnarray}
\chi_{abcd}(T) & = & \chi_{badc}^{*}(T),\label{eq:hermiticity}\\
\sum_{a}\chi_{aacd}(T) & = & \delta_{cd},\label{eq:trace preservation}\\
\sum_{abcd}z_{ac}^{*}\chi_{abcd}(T)z_{bd} & \geq & 0,\label{eq:positive semidefinite}
\end{eqnarray}
where $z$ is any complex valued vector. Using Eqs.\ \ref{eq:hermiticity}
and \ref{eq:trace preservation}, for a system in a $d$-dimensional
Hilbert space, $\chi(T)$ is determined by $d^{4}-d^{2}$ real valued
parameters \cite{nielsenchuangbook}. Operationally, QPT can be defined
as an experimental protocol to obtain $\chi(T)$. Spectroscopically,
$\chi(T)$ may be reconstructed by measuring $\rho(T)$ (i.e., performing
QST) given some choice of initial state $\rho(0)$, where the $\rho(0)$
are chosen successively from a complete set of initial states \cite{cory,opticallattices,natphysqpt,shabaniprl}.
Since we are interested in energy transfer dynamics, this procedure
shall be performed at several values of $T$. In this article, we
show how to perform QPT for a model coupled excitonic dimer using
two-color heterodyne photon-echo experiments.

\section{Description of the system and its interaction with light}

Consider an excitonic dimer interacting with a bath of phonons. The
excitonic part of the Hamiltonian, describing the system with the
environment frozen in place, is given by \cite{moran,minhaengbook,pullerits-prb}:

\begin{align}
H_{S} & =\omega_{A}a_{A}^{+}a_{A}+\omega_{B}a_{B}^{+}a_{B}+J(a_{A}^{+}a_{B}+a_{B}^{+}a_{A})\nonumber \\
 & =\omega_{\alpha}c_{\alpha}^{+}c_{\alpha}+\omega_{\beta}c_{\beta}^{+}c_{\beta}.\label{eq:excitonic H}\end{align}
where $a_{i}^{+}$ and $c_{j}^{+}$ ($a_{i}$ and $c_{j}$) are creation
(annihilation) operators for site $i\in\{A,B\}$ and delocalized $j\in\{\alpha,\beta\}$
excitons, respectively. $\omega_{A}\neq\omega_{B}$ are the first
and second site energies, $J\neq0$ is the Coulombic coupling between
the chromophores. We define the average of site energies $\omega=\frac{1}{2}(\omega_{A}+\omega{}_{B})$,
difference $\Delta=\frac{1}{2}(\omega_{A}-\omega_{B})$, and mixing
angle $\theta=\frac{1}{2}\arctan\left(\frac{J}{\Delta}\right)$. Then
$c_{\alpha}=\cos\theta a_{A}+\sin\theta a_{B}$, $c_{\beta}=-\sin\theta a_{A}+\cos\theta a_{B}$,
$\omega_{\alpha}=\omega+\Delta\sec2\theta$, and $\omega_{\beta}=\omega-\Delta\sec2\theta$.
For convenience, we define the single exciton states $|\alpha\rangle=c_{\alpha}^{+}|g\rangle$,
$|\beta\rangle=c_{\beta}^{+}|g\rangle$, where $|g\rangle$ is the
ground state, and the biexcitonic state $|f\rangle=a_{A}^{+}a_{B}^{+}|g\rangle=c_{\alpha}^{+}c_{\beta}^{+}|g\rangle$.
The model Hamiltonian does not account for exciton-exciton binding
or repulsion terms, so the energy level of the biexciton is $\omega_{f}=\omega_{\alpha}+\omega_{\beta}=\omega_{A}+\omega_{B}$.
Denoting $\omega_{ij}\equiv\omega_{i}-\omega_{j}$, we have $\omega_{\alpha g}=\omega_{f\beta}$
and $\omega_{\beta g}=\omega_{f\alpha}$.

We are interested in the perturbation of the excitonic system due to three laser pulses:

\begin{equation}
V(t')=-\lambda\sum_{i=1}^{3}\hat{\boldsymbol{\mu}}\cdot\boldsymbol{e_{i}}E(t'-t_{i}) \{e{}^{i\boldsymbol{k}_{i}\cdot\boldsymbol{r}-i\omega_{i}(t'-t_{i})}+c.c.\},\label{eq:perturbation}
\end{equation}
where $\lambda$ is the intensity of the electric field, assumed weak, $\hat{\boldsymbol{\mu}}$ is the dipole operator, and $\boldsymbol{e_{i}},t_{i},\boldsymbol{k}_{i},\omega_{i}$
denote the polarization,%
\footnote{We use the word \emph{polarization} in two different ways: To denote
(a) the orientation of oscillations of the electric field and (b)
the density of electric dipole moments in a material. The meaning
should be clear by the context. %
} time center, wavevector, and carrier frequency of the $i$-th pulse.
$E(t')$ is the slowly varying pulse envelope, which we choose to
be Gaussian with fixed width $\sigma$ for all pulses, $E(t')=e^{-t'^{2}/2\sigma^{2}}$.
The polarization induced by the pulses on the molecule located at
position $\boldsymbol{r}$ is given by $\boldsymbol{P}(\boldsymbol{r},t')=\trace(\hat{\boldsymbol{\mu}}\rho(\boldsymbol{r},t'))$.
This quantity can be Fourier decomposed along different wavevectors
as $\boldsymbol{P}(\boldsymbol{r},t')=\sum_{s}\boldsymbol{P}_{s}(t')e^{i\boldsymbol{k}_{s}\cdot\boldsymbol{r}}$,
where the $\boldsymbol{k}_{s}$ are linear combinations of wavevectors
of the incoming fields. Radiation is produced due to the polarization
(proportional to $i\boldsymbol{P}(\boldsymbol{r},t')$). We can choose
to study a single component $\boldsymbol{P}_{s}$ by detecting only
the radiation moving in the direction $\boldsymbol{k}_{s}$. This
is achieved by interfering the radiation with a fourth pulse moving
in the direction $\boldsymbol{k}_{s}$, called the local oscillator
(LO) \cite{mukamel}. In particular, we shall be interested in the
time-integrated signal in the photon-echo (PE) direction, $\boldsymbol{k}_{4}=\boldsymbol{k}_{PE}=-\boldsymbol{k}_{1}+\boldsymbol{k}_{2}+\boldsymbol{k_{3}}$.
This heterodyne-detected signal $[S_{PE}]{}_{\boldsymbol{e}_{1},\boldsymbol{e}_{2},\boldsymbol{e}_{3},\boldsymbol{e}_{4}}^{\omega_{1},\omega_{2},\omega_{3},\omega_{4}}$,
where the subscripts indicate the polarizations of the four light
pulses and the superscripts indicate their carrier frequencies, is proportional to
\begin{eqnarray}
\int_{-\infty}^{\infty}dt'e^{i\omega_{4}(t'-t_{4})}E(t'-t_{4})\boldsymbol{e_{4}}\cdot i\boldsymbol{P}_{PE}(t').\label{eq:signal}\end{eqnarray}
Spatial integration over the volume of probed molecules selects out the component
$\boldsymbol{P}_{PE}(t')$ from the $\boldsymbol{P}(\boldsymbol{r},t')$.
The time integration yields a signal that is proportional to the components
of $\boldsymbol{P}_{PE}(t')\cdot\boldsymbol{e}_{4}$ oscillating at
the frequencies of the LO, which is centered about $\omega_{4}$ %
\footnote{More precisely, the monitored signal is proportional to $\int_{-\infty}^{\infty}dte^{i[\omega_{4}(t'-t_{4})+\varphi]}E(t'-t_{4})\boldsymbol{e_{4}}\cdot i\boldsymbol{P}_{PE}(t')+c.c.$,
where two experiments are conducted by varying the phase $\varphi$
of the LO with respect to the emitted polarization to extract the
real and imaginary terms of Eq.\ \ref{eq:signal}. For purposes of
our discussion, it is enough to consider the complex valued signal.%
}. In this excitonic model, the only optically allowed transitions
are between states differing by one excitation, so the only nonzero
matrix elements of $\hat{\boldsymbol{\mu}}$ are $\mathbb{\boldsymbol{\mu}}_{ij}=\boldsymbol{\mu}_{ji}$
for $ij=\alpha g,\beta g,f\alpha,f\beta$ (SI-II).

\begin{figure}
  \includegraphics[width=\figwidth]{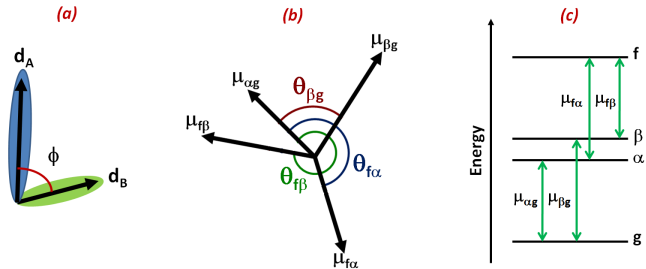}
  \caption{\emph{A set of photon echo experiments can be regarded as a QPT.}
Pulses are centered about $t_{1},t_{2},t_{3},t_{4}$. Time flows upwards
in the diagram. The four pulses define the \emph{coherence} ($\tau$),
\emph{waiting} ($T$) times, and echo ($t$) times. This experiment,
in the language of quantum information processing, can be regarded
as consisting of three stages: initial state preparation, free evolution,
and detection of the output state of the waiting time.}
\end{figure}

In the following section, we present the main results of our study.
We show that a carefully chosen set of two-color rephasing PE experiments
can be used to perform a \emph{QPT of the first exciton manifold}
(Fig.\ 1). The preparation of initial states is achieved using the
first two pulses at $t_{1}$ and $t_{2}$. Initial states spanning
the single exciton manifold are produced by using the four possible
combinations of two different carrier frequencies for the first two
pulses. In the terminology of PE experiments, these pulses define
the so-called \emph{coherence time} interval $\tau=t_{2}-t_{1}$.
The time interval between the second and third pulses, called the
\emph{waiting time} $T=t_{3}-t_{2}$, defines the quantum channel
\cite{nielsenchuangbook} which we want to characterize by QPT. Finally,
we carry out QST of the output density matrix at the instant $t_{3}$.
This task is indirectly performed by using the third pulse to selectively
generate new dipole-active coherences, which are detected upon heterodyning
with the fourth pulse at $t_{4}$, that is, after the \emph{echo time}
$t=t_{4}-t_{3}$ has elapsed. Varying the third and fourth pulse frequencies
yields sufficient linear equations for QST. This procedure naturally
concludes the protocol of the desired QPT.

\section*{Results}

For purposes of the QPT protocol, we assume that the structural parameters
$\omega_{\alpha g}$, $\omega_{\beta g}$, $\mathbb{\boldsymbol{\mu}}_{\alpha g}$,
$\mathbb{\boldsymbol{\mu}}_{\beta g}$, $\mathbb{\boldsymbol{\mu}}_{f\alpha}$,
and $\mathbb{\boldsymbol{\mu}}_{f\beta}$ are all known. Information
about the transition frequencies can be obtained from a linear absorption
spectrum, whereas the dipoles can be extracted from x-ray crystallography
\cite{spinach}. As shown in recent work of our group, with enough
data from the PE experiments, it is also possible to extract these
parameters self-consistently \cite{yuen-aspuru,marcus}. We proceed
to describe the steps of the PE experiment on a coupled dimer that
yield a QPT.

\subsection*{Initial state preparation}

Before any electromagnetic perturbation, the excitonic system is in
the ground state $\rho(-\infty)=|g\rangle\langle g|$. After the
first two pulses in the $\boldsymbol{k}_{1}$, $\boldsymbol{k}_{2}$ directions act on the system, the
effective density matrix $\tilde{\rho}_{\boldsymbol{e}_{1},\boldsymbol{e}_{2}}^{\omega_{1},\omega_{2}}(0)$
(at $T=0$) is created. This density is second order in $\lambda$ and, combined with the third and fourth pulses, directly determines the signal. 
By applying second
order perturbation theory and the rotating-wave approximation (RWA),
we can define an effective \emph{initial state} (Fig.\ 2a--d, SI-III):
\begin{eqnarray}
\tilde{\rho}_{\boldsymbol{e}_{1},\boldsymbol{e}_{2}}^{\omega_{1},\omega_{2}}(0) & = & -\sum_{pq\in\{\alpha,\beta\}}C_{\omega_{1}}^{p}C_{\omega_{2}}^{q}(\boldsymbol{\mu}_{pg}\cdot\boldsymbol{e}_{1})(\boldsymbol{\mu}_{qg}\cdot\boldsymbol{e}_{2})\nonumber \\
 &  & \times\mathcal{G}_{gp}(\tau)(|q\rangle\langle p|-\delta_{pq}|g\rangle\langle g|).\label{eq:initial state}\end{eqnarray}
 This state evolves during the waiting time $T$ to give \begin{eqnarray}
\tilde{\rho}_{\boldsymbol{e}_{1},\boldsymbol{e}_{2}}^{\omega_{1},\omega_{2}}(T) & = & \chi(T)\tilde{\rho}_{\boldsymbol{e}_{1},\boldsymbol{e}_{2}}^{\omega_{1},\omega_{2}}(0),\label{eq:second order}\end{eqnarray}
which holds for $T\gtrsim3\sigma$, that is, after the action of the
first two pulses has effectively ended. Eq.\ \ref{eq:second order}
is of the form of Eq.\ \ref{eq:linear transformation}, and therefore
appealing for our QPT purposes. The purely imaginary coefficients
$C_{\omega_{i}}^{p}$ for $p\in\{\alpha,\beta\}$ are proportional
to the frequency components at $\omega_{pg}$ of the pulse which is
centered at $\omega_{i}$:
\begin{equation}
C_{\omega_{i}}^{p}=-\frac{\lambda}{i}\sqrt{2\pi\sigma^{2}}e^{-\sigma^{2}(\omega_{pg}-\omega_{i})^{2}/2},\label{eq:coefficient}\end{equation}
 and the propagator of the optical coherence $|i\rangle\langle j|$ is
\begin{equation}
\mathcal{G}_{ij}(\tau)=\Theta(\tau)e^{(-i\omega_{ij}-\Gamma_{ij})\tau},\label{eq:optical coherence}
\end{equation}
which, for simplicity, has been taken to be the product of a coherent
oscillatory term beating at a frequency $\omega_{ij}$ and an exponential
decay with dephasing rate $\Gamma_{ij}$, assumed to be known. $\Theta(\tau)$
is the Heaviside function, so the propagator is finite only for times
$\tau\geq0$. We have kept only the $-\boldsymbol{k}_{1}$ and $+\boldsymbol{k}_{2}$
components because those are the only contributions to the signal
at $\boldsymbol{k}_{PE}$.

\begin{figure}
  \includegraphics[width=\figwidth]{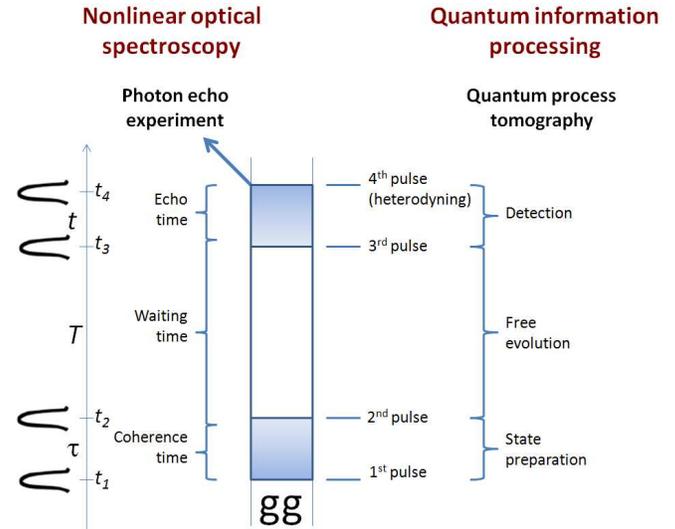}
  \caption{\emph{Possible state preparations and detections. }We list all the
possible preparations and detections of elements of the density matrix
at the waiting time $T$ via a rephasing PE experiment. The double
sided Feynman diagrams above list all the possible processes detected
in a rephasing PE experiment. Each diagram is related to an element
$\chi_{ijqp}(T)$ due to the prepared state $|q\rangle\langle p|-\delta_{pq}|g\rangle\langle g|$
at the beginning of the waiting time and the detected state $ij$
at the end of it. By combining the preparations in (a), (b), (c),
and (d), with the detections in (e), (f), (g), and (h), sixteen different
types of processes can be enumerated, which can be classified according
to the pulse frequencies of the associated perturbations.}
\end{figure}

Eq.\ (\ref{eq:initial state}) has a simple interpretation, and can
be easily read off from the Feynman diagrams depicted in Fig.\ 2a--d.
Since we will be selecting only light in the $\boldsymbol{k}_{PE}$
direction, we keep only the portion of the first pulse proportional
to $e^{i\omega_{1}(t-t_{1})}$ and only the portion of the second
pulse proportional to $e^{-i\omega_{2}(t-t_{2})}$ in Eq.\ \ref{eq:perturbation}.
Then, since the state before any perturbation is $|g\rangle\langle g|$,
the first pulse can only resonantly excite the bra in the RWA \cite{mukamel},
generating an optical coherence $|g\rangle\langle p|$ (where $p\in\{\alpha,\beta\}$)
with amplitude $C_{\omega_{1}}^{p}$. This coherence undergoes free
evolution for time $\tau$ under $\mathcal{G}_{gp}(\tau)$ before
the second pulse perturbs the system. In the RWA, this second pulse
can act on the ket of $|g\rangle\langle p|$ to yield $|q\rangle\langle p|$
with amplitude $C_{\omega_{2}}^{q}$ and on the bra to create a \emph{hole}
$-|g\rangle\langle g|$ with amplitude $C_{\omega_{2}}^{p}$, producing
Eq.\ \ref{eq:initial state}. The amplitude of this prepared initial
state is proportional to the alignment of the corresponding transition
dipole moments with the polarization of the incoming fields. Once
the initial state is prepared, it evolves via $\chi(T)$, which is
the object we want to characterize. A possible problem is the contamination
of the initial states by terms proportional to a hole$-|g\rangle\langle g|$
every time there is a single-exciton manifold population $|p\rangle\langle p|$.
This is not a difficulty if we assume:\begin{equation}
\chi_{abgg}(T)=\delta_{ag}\delta_{bg},\label{eq:xgg}\end{equation}
that is, if the ground state population does not transform into any
other state via free evolution. This is reasonable since we may ignore
processes where phonons can induce upward optical transitions. We
similarly neglect spontaneous excitation from the single to double
exciton states.

One can see from Eq.\ \ref{eq:initial state} that a set of four
linearly independent initial states can be generated by manipulating
the frequency components of the pulses through $C_{\omega_{1}}^{p}$
and $C_{\omega_{2}}^{q}$. It is sufficient to consider a pulse toolbox
of two waveforms which create $|\alpha\rangle$ and $|\beta\rangle$
with different amplitudes. For instance, by centering one waveform
at $\omega_{+}$ in the vicinity of $\omega_{\alpha g}$ and the other
at $\omega_{-}$ close to $\omega_{\beta g}$ , we can simultaneously
have: \begin{eqnarray}
C_{\omega_{+}}^{\alpha} & = & C_{\omega_{-}}^{\beta}=C',\nonumber \\
C_{\omega_{+}}^{\beta} & = & C_{\omega_{-}}^{\alpha}=C'',\label{eq:pulse toolbox}\end{eqnarray}
for purely imaginary numbers $C'\neq C''$. The conceptually simplest
choice, which we shall denote the \emph{maximum discrimination choice}
(MDC), and which is best for QPT purposes, is two waveforms each resonant
with only one transition, so $C'\gg C''$, and we can neglect $C''$.
Four linearly independent initial states can be prepared by choosing
the waveform of each of the first two pulses from this pulse toolbox.

\subsection*{Evolution}

The system evolves during the waiting time $T$ after the initial
state is prepared. Transfers between coherences and populations are
systematically described by $\chi(T)$. The components of $\tilde{\rho}_{\boldsymbol{e}_{1},\boldsymbol{e}_{2}}^{\omega_{1},\omega_{2}}$
evolve in time, described by $\chi(T)$, without assuming any particular
model for the bath or system-bath interaction, other than Eq.\ \ref{eq:xgg}.
By definition, the amplitude of the $ij$ component of $\tilde{\rho}_{\boldsymbol{e}_{1},\boldsymbol{e}_{2}}^{\omega_{1},\omega_{2}}(T)$
is proportional to $\chi_{ijqp}(T)$, with the exception of the $gg$
component, which is proportional to $\chi_{ggpq}(T)+\delta_{pq}$
due to the contamination of the hole in the initial state. We note
that de-excitation transfers from the single-exciton manifold to elements
involving the ground state ($gg,g\alpha,g\beta$) are expected to
be small, since such processes are also unlikely to occur on femtosecond
timescales due to either the phonon or the photon bath.
Detailed analysis shows that our protocol can detect decay into $gg$ but not into $g \alpha$ or $g \beta$.
We shall keep these $\chi_{ggqp}(T)$ terms in our equations
in order to monitor \emph{amplitude leakage errors} from the single-exciton
manifold, providing a consistency check for treating the single-exciton
manifold as an effective TLS in the timescale of interest.


\subsection*{Detection}

The last two pulses provide an indirect QST of the state after the
waiting time. The third perturbation along $+\boldsymbol{k}_{3}$
and centered at time $t_{3}$ will selectively probe certain coherences
and populations of $\tilde{\rho}_{\boldsymbol{e}_{1},\boldsymbol{e}_{2}}^{\omega_{1},\omega_{2}}(T)$.
As an illustration (see Fig.\ 2e,f), the component of the pulse matching
the transition energy $\omega_{\alpha g}=\omega_{f\beta}$ and proportional
to $C_{\omega_{3}}^{\alpha}$will, in the RWA, promote the resonant
transitions $|g\rangle\to|\alpha\rangle$ and $|\beta\rangle\to|f\rangle$
on the ket side, and the conjugate resonant transitions $\langle\alpha|\rightarrow\langle g|$
and $\langle f|\rightarrow\langle\beta|$ on the bra side, the latter
of which are irrelevant as we are ignoring transfers to the biexciton
state during the waiting time. These transitions will generate two
sets of optically active coherences in the echo interval: $\alpha g,\, f\beta$
which oscillate at frequency $\omega_{\alpha g}$, and $\beta g,\, f\alpha$
which oscillate at $\omega_{\beta g}$. These sets generate a polarization
which interferes with the LO yielding signals proportional to $C_{\omega_{4}}^{\alpha}$
and $C_{\omega_{4}}^{\beta}$, respectively. The propagator for the
echo time is taken to be as in Eq.\ \ref{eq:optical coherence}. 
A similar analysis can be repeated for the $C_{\omega_{3}}^{\beta}$
term. Analogously to the preparation stage, the same toolbox of two
different waveforms for the third and the fourth pulses allows discrimination
of all final states of $\tilde{\rho}_{\boldsymbol{e}_{1},\boldsymbol{e}_{2}}^{\omega_{1},\omega_{2}}(T)$.
Fig.\ 2 depicts double-sided Feynman diagrams for all possible combinations
of preparations and detections with four pulses each chosen from two
waveforms, yielding sixteen experiments. By keeping track of these
processes, the signal $[S_{PE}]{}_{\boldsymbol{e}_{1},\boldsymbol{e}_{2},\boldsymbol{e}_{3},\boldsymbol{e}_{4}}^{\omega_{1},\omega_{2},\omega_{3},\omega_{4}}$
may be compactly written as (SI-IV):\begin{eqnarray}
 &  & [S_{PE}]{}_{\boldsymbol{e}_{1},\boldsymbol{e}_{2},\boldsymbol{e}_{3},\boldsymbol{e}_{4}}^{\omega_{1},\omega_{2},\omega_{3},\omega_{4}}(\tau,T,t)\nonumber \\
 & \propto & \sum_{p,q,r,s}C_{\omega_{1}}^{p}C_{\omega_{2}}^{q}C_{\omega_{3}}^{r}C_{\omega_{4}}^{s}P{}_{\boldsymbol{e}_{1},\boldsymbol{e}_{2},\boldsymbol{e}_{3},\boldsymbol{e}_{4}}^{p,q,r,s}(\tau,T,t),\label{eq:total polarization}\end{eqnarray}
where the proportionality constant is purely real, and the expression
holds for $T,t>3\sigma$. The terms $P{}_{\boldsymbol{e}_{1},\boldsymbol{e}_{2},\boldsymbol{e}_{3},\boldsymbol{e}_{4}}^{p,q,r,s}(\tau,T,t)$
are loosely polarizations (in fact, they are proportional to $i$
times polarizations) %
\footnote{
By comparing Eqs.\ \ref{eq:signal} and \ref{eq:total polarization},
we notice that both $[S_{PE}]{}_{\boldsymbol{e}_{1},\boldsymbol{e}_{2},\boldsymbol{e}_{3},\boldsymbol{e}_{4}}^{\omega_{1},\omega_{2},\omega_{3},\omega_{4}}(\tau,T,t)$
and $P{}_{\boldsymbol{e}_{1},\boldsymbol{e}_{2},\boldsymbol{e}_{3},\boldsymbol{e}_{4}}^{p,q,r,s}(\tau,T,t)$
are related to $i\boldsymbol{P}_{PE}$  via a real proportionality
constant. Although we shall denote $P{}_{\boldsymbol{e}_{1},\boldsymbol{e}_{2},\boldsymbol{e}_{3},\boldsymbol{e}_{4}}^{p,q,r,s}(\tau,T,t)$
loosely as a polarization, when referring to its real and imaginary
parts, we must remember that they are proportional to the real and
imaginary parts of the signal $[S_{PE}]{}_{\boldsymbol{e}_{1},\boldsymbol{e}_{2},\boldsymbol{e}_{3},\boldsymbol{e}_{4}}^{\omega_{1},\omega_{2},\omega_{3},\omega_{4}}(\tau,T,t)$,
and to the imaginary and real parts of the actual polarization $\boldsymbol{P}_{PE}$,
respectively. %
} and are given by\begin{eqnarray}
 &  & P{}_{\boldsymbol{e}_{1},\boldsymbol{e}_{2},\boldsymbol{e}_{3},\boldsymbol{e}_{4}}^{p,q,\alpha,\alpha}(\tau,T,t)\nonumber \\
 & = & -(\boldsymbol{\mu}_{pg}\cdot\boldsymbol{e}_{1})(\boldsymbol{\mu}_{qg}\cdot\boldsymbol{e}_{2})\mathcal{G}_{gp}(\tau)\nonumber \\
 &  & \times\{[(\boldsymbol{\mu}_{\alpha g}\cdot\boldsymbol{e}_{3})(\boldsymbol{\mu}_{\alpha g}\cdot\boldsymbol{e}_{4})\mathcal{G}_{\alpha g}(t)\nonumber \\
 &  & \times(\chi_{ggqp}(T)-\delta_{pq}-\chi_{\alpha\alpha qp}(T))\nonumber \\
 &  & +(\boldsymbol{\mu}_{f\beta}\cdot\boldsymbol{e}_{3})(\boldsymbol{\mu}_{f\beta}\cdot\boldsymbol{e}_{4})\mathcal{G}_{f\beta}(t)\chi_{\beta\beta qp}(T)]\},\label{eq:pqaa}\end{eqnarray}
 and

\begin{eqnarray}
 &  & P{}_{\boldsymbol{e}_{1},\boldsymbol{e}_{2},\boldsymbol{e}_{3},\boldsymbol{e}_{4}}^{p,q,\alpha,\beta}(\tau,T,t)\nonumber \\
 & = & -(\boldsymbol{\mu}_{pg}\cdot\boldsymbol{e}_{1})(\boldsymbol{\mu}_{qg}\cdot\boldsymbol{e}_{2})\mathcal{G}_{gp}(\tau)\nonumber \\
 &  & \times\{[((\boldsymbol{\mu}_{f\beta}\cdot\boldsymbol{e}_{3})(\boldsymbol{\mu}_{f\alpha}\cdot\boldsymbol{e}_{4})\mathcal{G}_{f\alpha}(t)\nonumber \\
 &  & -(\boldsymbol{\mu}_{\alpha g}\cdot\boldsymbol{e}_{3})(\boldsymbol{\mu}_{\beta g}\cdot\boldsymbol{e}_{4})\mathcal{G}_{\beta g}(t))\chi_{\beta\alpha qp}(T)]\}.\label{eq:pqab}\end{eqnarray}
 The remaining terms $P{}_{\boldsymbol{e}_{1},\boldsymbol{e}_{2},\boldsymbol{e}_{3},\boldsymbol{e}_{4}}^{p,q,\beta,\beta}(\tau,T,t)$,
$P{}_{\boldsymbol{e}_{1},\boldsymbol{e}_{2},\boldsymbol{e}_{3},\boldsymbol{e}_{4}}^{p,q,\beta,\alpha}(\tau,T,t)$
follow upon the interchange $\alpha\leftrightarrow\beta$. Eqs.\ \ref{eq:total polarization}-\ref{eq:pqab}
are the main result of this article. Each $P{}_{\boldsymbol{e}_{1},\boldsymbol{e}_{2},\boldsymbol{e}_{3},\boldsymbol{e}_{4}}^{p,q,r,s}$
represents the observed signal if the first (second,third,fourth)
laser pulse is resonant only with the $p$ ($q$,$r$,$s$) transition.
The total measured signal $[S_{PE}]{}_{\boldsymbol{e}_{1},\boldsymbol{e}_{2},\boldsymbol{e}_{3},\boldsymbol{e}_{4}}^{\omega_{1},\omega_{2},\omega_{3},\omega_{4}}(\tau,T,t)$
is a weighted sum of these $P{}_{\boldsymbol{e}_{1},\boldsymbol{e}_{2},\boldsymbol{e}_{3},\boldsymbol{e}_{4}}^{p,q,r,s}$.
Equations \ref{eq:pqaa},\ref{eq:pqab} show that each $P{}_{\boldsymbol{e}_{1},\boldsymbol{e}_{2},\boldsymbol{e}_{3},\boldsymbol{e}_{4}}^{p,q,r,s}$
is a linear combination of elements of $\chi(T)$, corresponding to
the prepared initial states and detected final states. After collecting
the sixteen signals $[S_{PE}]{}_{\boldsymbol{e}_{1},\boldsymbol{e}_{2},\boldsymbol{e}_{3},\boldsymbol{e}_{4}}^{\omega_{1},\omega_{2},\omega_{3},\omega_{4}}$
with each pulse carrier frequency $\omega_{i}$ chosen from $\{\omega_{+},\omega_{-}\}$
as in Eq.\ (\ref{eq:pulse toolbox}), with fixed polarizations $\boldsymbol{e}_{i}$,
Eq.\ (\ref{eq:total polarization}) can be inverted to yield the
elements of $\chi(T)$ associated with the single-exciton manifold
of the dimer, hence accomplishing the desired QSTs and QPT at once (see Table 1).
Notice that in principle, for a given value of waiting time $T$,
the one-dimensional (1D) measurements associated with a single set
of $\tau,t$ values is enough for purposes of QPT of the single exciton
manifold. In the most typical measurements, the sample has isotropically
distributed chromophores, so Eq.\ \ref{eq:total polarization} must
be modified to include isotropic averaging $\langle\cdot\rangle_{\text{iso}}$
(SI-VI). Since the present QPT protocol does not rely on different
polarization settings, we will assume for simplicity that each of
the sixteen experiments is carried out with $(\boldsymbol{e}_{1},\boldsymbol{e}_{2},\boldsymbol{e}_{3},\boldsymbol{e}_{4})=(z,z,z,z)$.
Further technical details of the QPT protocol can be found in the
next section as well as in SI-V--XII.

\subsection*{Important observations}

\newcounter{obs}\global\long\def\theobs{\alph{obs}}
\global\long\def\Obs{\refstepcounter{obs}(\theobs)}
 \global\long\def\noun#1{\textsc{#1}}

\Obs \noun{ }\textbf{Difference between a standard PE experiment
and QPT.}\noun{ } In the current practice of NLOS, model Hamiltonians
with free parameters for the excitonic system, the bath, and the interaction
between them are postulated, and the experimental spectra are fit
to the model via calculation of response functions, from which structural
and dynamical information is extracted \cite{khaliljpc}. In our language,
such experiments involve fitting models to complicated combinations
of quantum processes associated with $\chi(T)$. QPT requires only
a model for the excitonic system, but not for the bath or the system-bath
coupling, making it suitable for probing systems where the bath dynamics
are unknown. By definition, QPT extracts the elements of $\chi(T)$,
allowing a straightforward analysis of processes directly associated
with the density matrix, such as dephasing and relaxation.

\Obs \noun{ }\textbf{QPT can also be performed with control of time
delays $\tau$, $t$ instead of frequency control.}\noun{ } Although
1D measurements suffice for QPT, suppose the signal is collected for
many values of $\tau$ and $t$. Upon appropriately defined Fourier
transformations of the signal along these variables, a two-dimensional
electronic spectrum (2D-ES) can be constructed where the coherence
propagators of Eq.\ (\ref{eq:optical coherence}) manifest as four
resonances about $\omega_{\alpha g}$ and $\omega_{\beta g}$ along
both axes \cite{minhaengbook,jonas,yuen-aspuru}. An important
observation follows: the frequencies of the coherent evolutions in
the coherence and echo times are the same as the frequencies of the
first transition and the LO detection. By varying $t$ and $\tau$,
a 2D-ES provides the frequency-controlled information of the first
and fourth pulses. Hence, it is possible to make the first and fourth
pulses sufficiently broadband that their frequency components at the $\alpha g$ and
$\beta g$ transitions are of similar magnitude. Then, the sixteen 1D experiments
can be replaced with four 2D-ES where the second and third pulses
are frequency controlled. A caveat in this identification is the assumption
that the optical coherences evolve in a form like Eq.\ \ref{eq:optical coherence},
without errors of coherence transfers (SI-VII,IX).

\Obs \noun{ }\textbf{Extension to overlapping pulses.}\noun{ } The
discussion above assumed negligible pulse overlaps. Remarkably, Eqs.\ (\ref{eq:total polarization}),
(\ref{eq:pqaa}) and (\ref{eq:pqab}) still hold in general for any
$\tau,t\geq0$ and $T>3\sigma$, with the exception that $C_{\omega_{3}}^{r}C_{\omega_{4}}^{s}$
in Eq.\ \ref{eq:total polarization} must be replaced by \begin{align}
C_{\omega_{3}}^{r}C_{\omega_{4}}^{s}\frac{1}{2}\left[1+Erf\left(\frac{t}{2\sigma}+\frac{i(\omega_{3}-\omega_{rg}+\omega_{4}-\omega_{sg})\sigma}{2}\right)\right]\label{eq:modification pulse overlap-1}\end{align}
to account for the fact that the third pulse must act in the sample
\emph{before }the LO can detect the polarization (SI-IV-B). In the
case of well-separated pulses, $t\gg\sigma$, Eq.\ \ref{eq:modification pulse overlap-1}
reduces to $C_{\omega_{3}}^{r}C_{\omega_{4}}^{s}$.

The measurements of the real and imaginary part of the PE signal in
the $\tau,t=0$ limit are recognized with the names of transient dichroism
(TD) and transient birefringence (TB), respectively \cite{cho-fleming-mukamel},
and are very interesting for QPT. For resonant TD/TB ($\omega_{3}+\omega_{4}=\omega_{rs}+\omega_{sg}$),
Eq.\ (\ref{eq:modification pulse overlap-1}) reduces to $\frac{1}{2}C_{\omega_{3}}^{r}C_{\omega_{4}}^{s}$,
which shows that the LO monitors only half of the original polarization
since it interferes with the polarization as it is generated. Consider
such a resonant TD/TB experiment where, even though the pulses can
achieve frequency selectivity, they are short in the sense that $\sigma\ll\frac{1}{\lambda}$,
where $\lambda$ is a characteristic reorganization energy scale of
the bath. In this situation, the bath state will not evolve during
the action of the first two pulses, allowing unambiguous preparation
of initial excitonic states tensored with the \emph{same }initial
equilibrium bath configuration (SI-VIII), yielding a consistent QPT.
Also, for $\tau,t=0$, the free evolution of the optical coherences
does not contribute to the signal, and the short timescale $\sigma$
does not allow for errors of population or coherence transfers to
occur in the preparation or detection stages. Hence, a highlight of
the TD/TB signal is that it is determined exclusively by the dynamics
of the single-exciton manifold. Scenarios where the TD/TB configuration
is preferred compared to the PE are excitonic systems coupled to highly
non-Markovian baths \cite{cinakilinhumble,cinabiggs1,cinabiggs2,geva}.

\Obs\textbf{ Numerical stability of QPT.}\noun{ }An investigation
of the stability properties of the matrices associated with the reconstruction
of $\chi(T)$ from the sixteen enumerated experiments shows that our
protocol is very robust upon the variation of the structural parameters
of the system, namely, the ratio between the two dipole norms $d_{B}/d_{A}$,
the angle between the site dipoles $\phi$, and the mixing angle $\theta$.
General exceptions occur at the vicinity of $\theta=0,\frac{\pi}{2}$
where the coupling $J$ vanishes, as well as for $\theta=\frac{\pi}{4},\frac{3\pi}{4}$
and $d_{B}/d_{A}=1$, that is, the homodimer case (SI-XI-A and B).

\subsection*{Numerical example}

To test the extraction of $\chi$ from experimental spectra, we consider
a dimer with Hamiltonian parameters that are on the order of previously
reported experiments consisting of light harvesting systems\emph{
}\cite{HohjaiLee06082007,pullerits-prb}\emph{ }($\omega_{A}=12881$
cm$^{-1}$, $\omega_{B}=12719$ cm$^{-1}$, $J=120$ cm$^{-1},$ yielding
$\theta=0.49$). We assume a toolbox of two carrier frequencies $\omega_{+}=13480$
cm$^{-1}$ and $\omega_{-}=12130$ cm$^{-1}$, respectively, so that
$\omega_{i}\in\{\omega_{+},\omega_{-}\}$ for all $i$, and the width
of the pulses to be $FWHM=28.3$ fs in intensity, which corresponds to $\sigma=40$
fs in amplitude. The parameters satisfy the MDC condition with $C'/C''=20$. The
pulses are long enough to guarantee the selectivity of the produced
exciton, but short enough to allow for the evolution of the bath induced
excitonic dynamics to be monitored. We choose $d_{B}/d_{A}=2$ and
$\phi=0.3$. We present simulations on the QPT for this system, where
each chromophore is linearly coupled to an independent Markovian bath
of harmonic oscillators. The dissipative effects are modeled through
a secular Redfield model at temperature $\underbar{T}=273$ K (SI-X).

Since we are working in the MDC regime, the signals in Eq.\ \ref{eq:total polarization}
are simply proportional to $\langle P{}_{\boldsymbol{e}_{1},\boldsymbol{e}_{2},\boldsymbol{e}_{3},\boldsymbol{e}_{4}}^{p,q,r,s}(\tau,T,t)\rangle_{iso}$.
Fig.\ 3 plots the sixteen real and imaginary parts of the $\langle P{}_{z,z,z,z}^{p,q,r,s}(0,T,0)\rangle_{iso}$
values, which can be regarded as signals from the TD/TB setting, or
as PE signals with the coherence and echo time propagators factored
out. They have been calculated via an isotropic average of Eqs.\ (\ref{eq:pqaa})
and (\ref{eq:pqab}) (SI-VI,XI-B). In our simulations, we consider
an inhomogeneously broadened ensemble of 10,000 dimers with diagonal
disorder. The site energies are drawn from Gaussian distributions
centered about $\omega_{A}$ and $\omega_{B}$, respectively, both
with standard deviation of $\sigma_{inh}=40$~cm$^{-1}$. For every
waiting time $T$, the signal is calculated with this fixed ensemble.
After a normalization step, the signals are of \emph{O}(1) or smaller.
Additional noise simulating experimental errors due to laser fluctuations
is included. This consists of independent realizations at every waiting
time $T$ of Gaussian noise on the measured signals with zero mean
and $\sigma_{laser}=0.05$ standard deviation.

Fig.\ 3 plots the ideal and inhomogeneously broadened, noisy signals
as continuous and discrete points, respectively. The ideal signals
are calculated from a single dimer with no disorder and without laser
fluctuations. All plots start at $T=3\sigma=51$ fs, since for earlier
times, the initial states are not yet effectively prepared. Errors
in experimental signals $\langle P{}_{z,z,z,z}^{p,q,r,s}(0,T,0)\rangle_{iso}$
translate into errors of reconstructed $\chi(T)$. An estimate of
the amplification of relative errors is set by the condition numbers
of the matrices to be inverted, which is lower than $\kappa=2.9$
for our set of parameters $d_{B}/d_{A}$, $\theta$, $\phi$ (SI-XI-B).
Reconstruction of $\chi$ must respect the known symmetries, Eqs.\ \ref{eq:hermiticity}-\ref{eq:positive semidefinite}.
Eqs.~\ref{eq:hermiticity}-\ref{eq:trace preservation} are built
into the corresponding matrix equations, but Eq.\ (\ref{eq:positive semidefinite})
must be included by using a semidefinite programming routine. The
latter is implemented using the open-source package CVX \cite{cvx},
and the result is Fig.\ 4, which shows the discrete points representing
the reconstructed elements of $\chi(T)$ from noisy data on top of
the ideal results plotted as continuous functions of $T$. The relative
error of the inverted $\chi(T)$ averages to 0.12. Notice that despite
the significant inhomogeneous broadening and noise,
there is remarkable agreement between the ideal and the reconstructed values.
This finding is reminiscent of studies due to Humble and Cina \cite{humblecinajpc}.

Fig.\ 4 illustrates the final objective of a coupled dimer QPT, namely,
the process matrix $\chi(T)$. Each panel shows processes of the density
matrix as a function of $T$, conditional on the initial state being
$\alpha\alpha$ (a), $\beta\beta$ (b), or $\alpha\beta$ (c and
d), with the ideal $T$-dependence. The detailed balance condition
in the Redfield model implies that $\chi_{\beta\beta\alpha\alpha}=\chi_{\alpha\alpha\beta\beta}e^{\omega_{\alpha\beta}/k_{B}\underbar{T}}$,
which can be seen in panels (a) and (b). Also, note that due to the
secular approximation, coherence-to-population and the reverse processes
are zero. Note that all the decoherence processes in our model occur
within a timescale of hundreds of femtoseconds, with the $\alpha\beta$
coherence evolving through about three periods before practically
vanishing (c and d).

Clearly, a more complex interplay of the excitonic
system with the phonon bath is possible \cite{leggett},
but this example illustrates the essence of the type of information
which can be obtained through QPT.

\section*{Conclusions}

In this article, we have introduced QPT as a powerful tool to systematically
characterize the dynamics of excitonic systems in condensed phases.
We identified the coherence, waiting, and echo intervals of the PE
experiment with the state preparation, free evolution, and detection
stages of a QPT. In order to achieve selective state preparation and
detection, we suggested frequency control through pulses of two different
colors, although scenarios with time delays and pulse polarizations
as control knobs were also discussed here and elsewhere \cite{procedia,yuen-aspuru}.
By choosing between these colors for each of the four pulses, sixteen
experiments can be carried out, which yield all the elements of $\chi(T)$
related to the single-exciton manifold. An analysis of the reconstruction
of $\chi(T)$ in the presence of inhomogeneous broadening and experimental
noise was provided, and the simulation on a model system shows that
QPT of an excitonic system in condensed phase is a very plausible
goal.

Equipped with $\chi(T)$, which completely characterizes the excitonic
dynamics, a plethora of questions can be rigorously answered about
it. Some examples are: Can the bath be described as Markovian? If
so, does the secular approximation hold, or can a population spontaneously
be transferred to a coherence? \cite{ishizakijcp} If not, what is
its degree of non-Markovianity? \cite{ChengEngel,patricknonmarkovian}
Does a given master equation accurately describe the dynamics of the
system? What is the timescale of each decoherence process? Are the
baths coupled to each chromophore correlated? \cite{AkihitoIshizaki10132009,
korotkov}
How much entanglement is induced in the system upon photoexcitation?
\cite{sarovar} Once these questions are answered, interesting questions
of control \cite{brumer} and manipulation of excitons can be asked.

In summary, a QIP approach to nonlinear spectroscopy via QPT offers
novel insights on the ways to design experiments in order to extract
information about the quantum state of the energy transfer system.
We believe this work bridges a gap between theoretical and experimental
studies on excitation energy transfer from the QIP and physical chemistry
communities, respectively.

\begin{figure}
  \includegraphics[width=\figwidth]{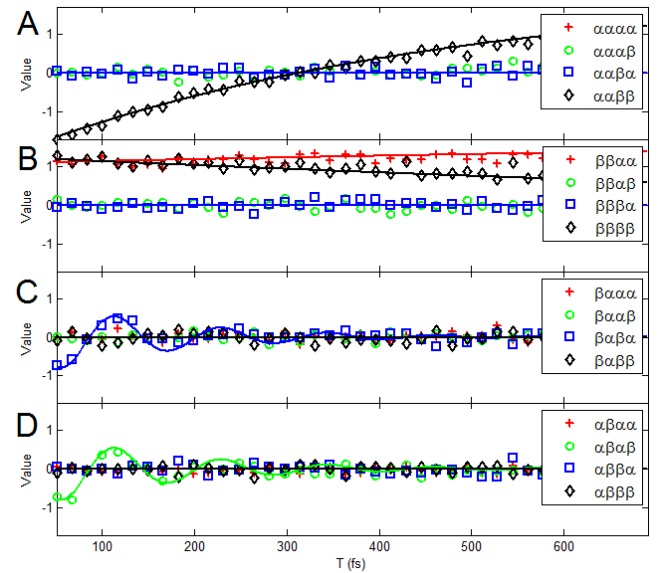}
  \caption{\emph{Polarization signals from sixteen two-color experiments for fixed
$\tau=t=0$. }The legends $pqrs$ correspond to the real parts of the isotropically averaged signals $\langle P{}_{z,z,z,z}^{p,q,r,s}(0,T,0)\rangle_{iso}$.
The panels are organized by QPT initial state: (a) for $|\alpha\rangle\langle\alpha|$,
(b) for $|\beta\rangle\langle\beta|$, (c) for $|\alpha\rangle\langle\beta|$,
and (d) for $|\beta\rangle\langle\alpha|$). The ideal signals
are depicted as continuous functions whereas the simulations with
inhomogeneous broadening and noise are represented as discrete points
of the same color.}
\end{figure}

\begin{figure}
  \includegraphics[width=\figwidth]{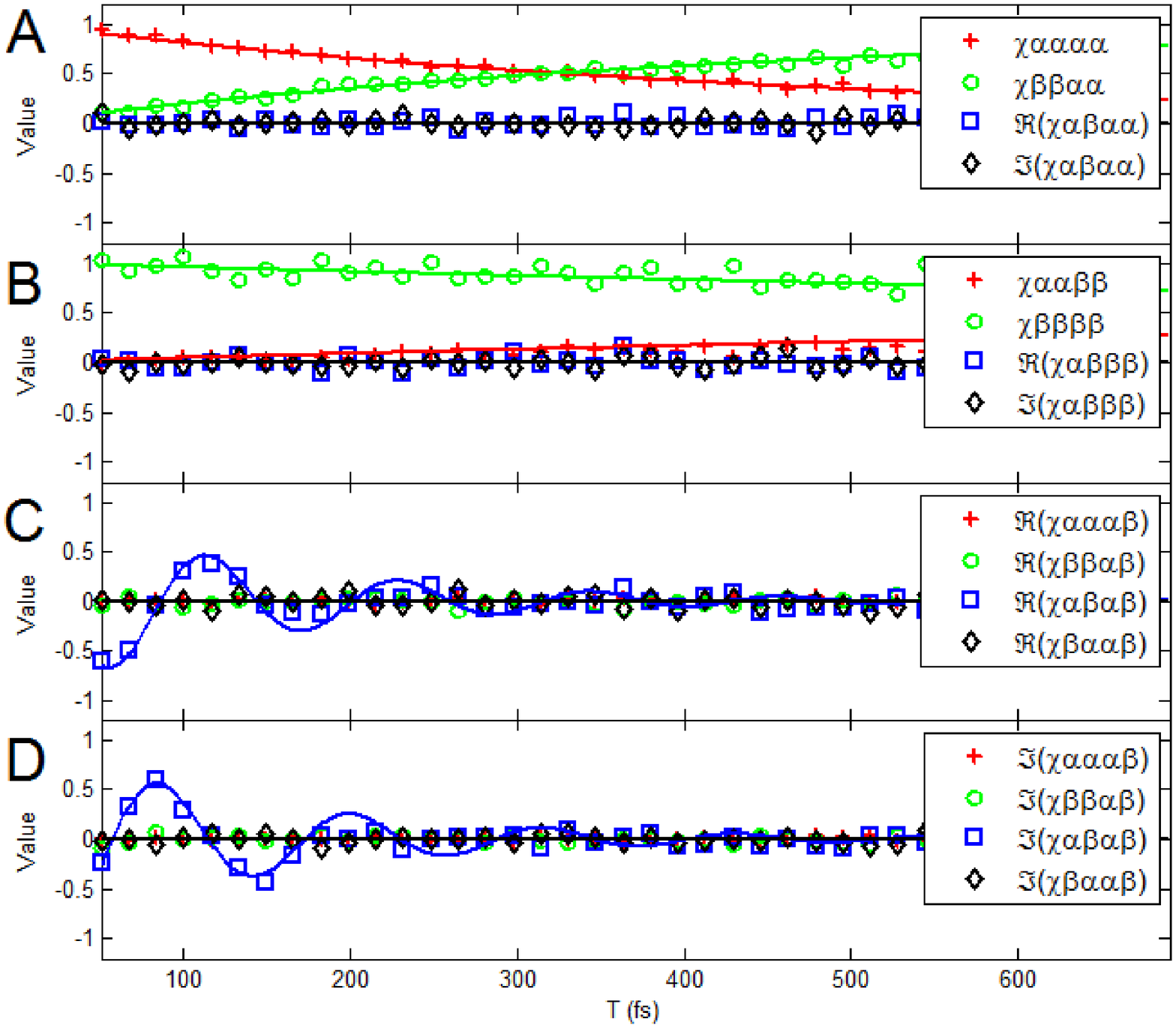}
  \caption{\emph{Elements of $\chi(T)$ for the numerical example.} The true
values are shown as a continuous function, whereas the discrete points
represent the extraction from noisy data. The panels are organized
according to the initial state, (a) for $|\alpha\rangle\langle\alpha|$,
(b) for $|\beta\rangle\langle\beta|$, and (c) and (d) for $|\alpha\rangle\langle\beta|$.}
\end{figure}

\begin{acknowledgments}
We acknowledge stimulating discussions with Hohjai Lee, Dylan Arias,
Patrick Wen, and Keith Nelson. This work was supported by the Center
of Excitonics, an Energy Frontier Research Center funded by the
US DOE, Office of Science, Office of Basic Energy
Sciences under Award Number DESC0001088 and the Harvard
University Center for the Environment.
\end{acknowledgments}

\bibliographystyle{pnas2009}

\end{article}

\end{document}